\documentclass[twocolumn,floatfix,superscriptaddress,amsmath,showpacs,showkeys,aps,prb]{revtex4-1}

\usepackage[utf8]{inputenc}
\usepackage[final]{graphicx}
\usepackage{epsfig}
\usepackage{bm}
\usepackage[normalem]{ulem}
\usepackage[usenames]{color}
\usepackage{times}

\newcommand{\beq}{\begin{equation}}
\newcommand{\eeq}{\end{equation}}
\newcommand{\beqa}{\begin{eqnarray}}
\newcommand{\eeqa}{\end{eqnarray}}
\newcommand{\vecx}{\vec{x}}
\newcommand{\veck}{\vec{k}}
\newcommand{\ddk}{\frac{d^2 k}{(2\pi)^2}}

\begin{document}

\title{Microscopic approach to orientational order of domain walls}

\author{Daniel G. Barci}
\affiliation{Departamento de F{\'\i}sica Te\'orica,
Universidade do Estado do Rio de Janeiro, Rua S\~ao Francisco Xavier 524, 20550-013  
Rio de Janeiro, Brazil}
\author{Daniel A. Stariolo}
\email{daniel.stariolo@ufrgs.br}
\affiliation{Departamento de F\'{\i}sica,
Universidade Federal do Rio Grande do Sul and\\
National Institute of Science and Technology for Complex Systems\\
CP 15051, 91501-970 Porto Alegre, RS, Brazil}

\date{\today}

\begin{abstract}
We develop a fully microscopic, statistical mechanics approach to study phase transitions
in Ising systems with competing interactions at different scales. Our aim is to consider orientational
and positional order parameters in a unified framework. In this work we consider two dimensional
stripe forming systems, where nematic, smectic and crystal phases are possible. We introduce a
nematic order parameter in a lattice, which measures orientational order of interfaces.
We develop a mean field approach which leads to a free energy which is a function of
both the magnetization (density) and the orientational (nematic) order parameters. Self-consistent
equations for the order parameters are obtained and the solutions are described for a
particular system, the Dipolar Frustrated Ising Ferromagnet. We show that this system
has an Ising-nematic phase at low temperatures in the square lattice, 
where positional order (staggered magnetization)
is zero. At lower temperatures a crystal-stripe phase may appear. In the continuum limit
the present approach connects to a Ginsburg-Landau theory, which has
an isotropic-nematic phase transition with breaking of a continuous symmetry. 
\end{abstract}

\pacs{64.60.De,75.70.Ak,75.30.Kz,75.70.Kw}

\keywords{}

\maketitle

\section{Introduction}
Systems with competing interactions are a rule in nature. Competing tendencies are usually
responsible for the complex behavior of natural systems, leading to slow dynamics,
metastability and complex free energy landscapes, like observed in spin glasses and
many other frustrated systems. Competing interactions at different scales may give
rise to complex phases and patterns, like stripes, lamellae, bubbles and others~\cite{SeAn1995}.
Examples range from solid state systems, like ultrathin ferromagnetic films
~\cite{PoVaPe2003,WoWuCh2005} and strongly correlated electron systems~\cite{KiFrEm1998,
FrKi1999}, to soft matter systems like Langmuir monolayers~\cite{SeMoGoWo1991},
block copolymers~\cite{VeHaAnTrHuChRe2005,RuBoBl2008},
colloids and soft core systems~\cite{MaPe2004,ImRe2006,GlGrKaKoSaZi2007}. Besides the
intrinsic interest raised by the complexity of the phase behavior in these systems, 
their detailed knowledge may be
critical for understanding basic phenomena as high temperature superconductivity, and
also for technological applications like soft matter templates for nanoscale systems and
future spintronic devices.

Systems with competing interactions at different scales, e.g. a short range attraction 
and a long range repulsion, as present in magnetic films and low dimensional electronic
systems, have a tendency to form {\em microphases}, i.e. to phase separate at
mesoscopic scales, due to the frustration usually present as a consequence of competing effects.
These microphases can show at least two types of ordering: positional ordering of the
microscopic degrees of freedom and also orientational order of interfaces or domain walls,
due to the presence of interfaces at mesoscopic scales. The presence of orientational order
allows for an analogy with liquid-crystal phenomenology, where nematic, smectic and crystal
phases can in principle be characterized. Nevertheless, a complete characterization of
the phases present in these systems is rare, and usually only positional order parameters
are computed and the corresponding phase diagrams are well known
~\cite{GaDo1982,AnBrJo1987,GrTaVi2000,PiCa2007,CaCaBiSt2011}.
In stripe forming systems, like the ones studied in the previous references, these mean field 
approaches give access only to modulated crystalline phases, or stripe-crystals, where both
positional and orientational long range order are simultaneously present. 
In fact, the morphologies of domains observed in real films, like the magnetization
patterns in ultrathin ferromagnetic films~\cite{VaStMaPiPoPe2000,PoVaPe2003,WoWuCh2005}
for example, give clear indications that orientational order sets in well before any
positional order can appear. It is necessary to go beyond the usual mean field approaches in
order to describe the complete ordering process. To our knowledge, there have been very few attempts 
to compute full phase diagrams, with simultaneous and independent consideration of orientational
and positional orders. One of these few works is the phase diagram for ultrathin ferromagnetic
films with square lattice anisotropy anticipated by Abanov and coworkers~\cite{AbKaPoSa1995}.
These authors have analyzed the phases of an Heisenberg spin system in the square lattice
with competing exchange and dipolar interactions, by means of a mixture of microscopic
calculation with phenomenological assumptions. Their analysis gave a very rich phase
diagram, with isotropic, Ising-nematic and smectic phases. Phenomenological approaches are
common ground for the analysis of orientational order. Starting from an elastic energy which
assumes a crystalline ground state, specific fluctuations of the ground state can be studied 
perturbatively. A complete analysis of two dimensional smectic elasticity
has been done thirty years ago by Toner and Nelson~\cite{ToNe1981}. 

Another interesting approach to study the
 phase transitions in systems with isotropic competing interactions was due to Brazovskii
~\cite{Br1975}. Analyzing a generic Ginsburg-Landau model whose main characteristic is the
presence of a minimum in the spectrum of Gaussian fluctuations at a non-zero wave vector,
he showed, in the context of a self-consistent field approximation,  that the model has a 
first order transition  to a modulated phase, induced by fluctuations. While conceptually 
important, this result suffers from the same problems
as other mean field approaches, namely, it describes phases where both orientational (modulated)
and positional order sets in simultaneously and has very strong transversal fluctuations in
two dimensions~\cite{SwHo1977}. In a recent work we have gone beyond the Brazovskii approximation,
by considering other terms in the Ginsburg-Landau free energy of two dimensional systems, 
which are dictated by the
symmetry of the interactions~\cite{BaSt2007,BaSt2009}. Interestingly, pure symmetry considerations
lead to terms in the free energy that encode orientational order parameters. We were
able to find a nematic phase, at temperatures above the original Brazovskii modulated phase.
By analyzing fluctuations of the mean field nematic solution we further found that the 
isotropic-nematic phase transition in the continuum system with isotropic interactions is of the
Kosterlitz-Thouless type~\cite{BaSt2009}. These works set the stage for
a complete Renormalization Group treatment of this kind of systems. Nevertheless, universality
in two dimensions is a delicate matter, and it would be important to get contact with the
phenomenological theory from more microscopic, interaction specific, models. 

In this work we give a step further in the description of nematic order in systems with
competing interactions, and develop a fully microscopic statistical mechanical approach 
which allows to consider positional as well as orientational order parameters on equal
grounds. We study Ising systems with arbitrary competing interactions on a square lattice.
We introduce a nematic order parameter on the lattice, by analogy with the order parameter
studied in the framework of the Ginsburg-Landau model~\cite{BaSt2009}. Then we develop a
mean field approach which includes both positional (magnetization) and orientational
(nematic) order parameters in the free energy. Due to the two-body nature of the orientational
order parameter it is necessary to go beyond one site approximations in order to compute
the mean field free energy. A set of self-consistent equations for the order parameters,
equivalent to the Bragg-Williams approximation for the Ising model, is obtained and the
solution for a model system presented. We computed the phase diagram of the Dipolar Frustrated
Ising Ferromagnet (DFIF), a well known and very studied model for ultrathin magnetic films with
perpendicular anisotropy~\cite{CzVi1989,DeMaWh2000,CaMiStTa2006,ViSaPoPePo2008}. We show
that this model has an Ising-nematic phase in the square lattice, where only orientational
order is present. At lower temperatures a further transition to a modulated phase with
positional order, the stripe phase, is possible. In the continuum limit our approach leads
to the Ginsburg-Landau model discussed before~\cite{BaSt2007}. It is also shown that,
when considering the nematic order parameter, the
magnetic (spin-spin) susceptibility diverges at a higher temperature compared with the
usual mean field calculation, in which the orientational terms in the free
energy are absent. 

The paper is organized as follows: in section \ref{OrderParameter} we introduce the nematic and 
the Ising-nematic order parameters. In this section we also develop a mean field approach to deal with 
general competing interactions in the square lattice. In \S \ref{dipolar} we apply our technique to the 
Dipolar Frustrated Ising Ferromagnet. Finally we discuss our results and conclusions in \S \ref{conclusions}.

\section{Microscopic Nematic-like order parameter}
\label{OrderParameter}
\begin{figure}
\begin{center}
\includegraphics[scale=0.8]{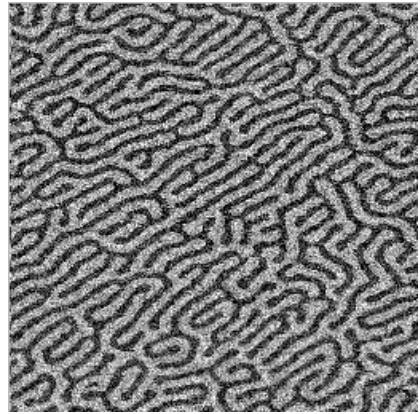}
\caption{Image of domain patterns in a Fe/Cu(001) ultrathin film, taken by Scanning
Electron Microscopy with Polarization Analysis (SEMPA). Courtesy of Prof. Danilo
Pescia, ETH, Zurich.}
\label{fig1}
\end{center}
\end{figure}

Figure \ref{fig1} shows a typical pattern of magnetic domains in an ultrathin 
ferromagnetic film of Fe/Cu(001) with strong perpendicular anisotropy. In this
system, the competing interactions are the short range exchange interaction
and the long range dipolar interaction. It is clear that this pattern does not
present positional order, but
it is possible to distinguish some degree of orientational order in the
stripe-like pattern. We want to define a suitable order parameter for this kind
of order and quantify it. 


Domain walls are observed at the transitions
between positive and negative values of the perpendicular magnetization. In an appropriate mesoscopic scale, 
it is possible to define a continuum magnetization density  $\Phi(\vecx)$. The gradient of this quantity 
\beq
\mathbf N(\vecx) \equiv \nabla \phi(\vecx) = (\partial_x\phi, \partial_y\phi)
\eeq
naturally defines a director that quantifies the degree of orientation of domain walls.

\begin{figure}[h]
\begin{center}
\includegraphics[scale=0.4]{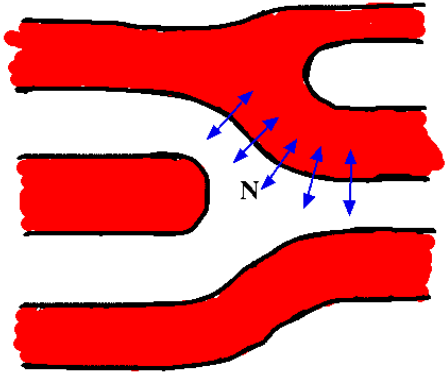}
\caption{(Color online) Schematic representation of a stripe phase with a dislocation. 
The arrows represent
the director field, which is a local measure of orientation of domain walls.}
\label{fig2}
\end{center}
\end{figure}

However,  the kind of order we are looking for is insensitive to the vector orientation, as shown in
figure \ref{fig2}. Therefore,  in analogy with the nematic order parameter of liquid crystals, 
it is possible to define a local tensor order parameter:
\beq
Q_{ij}(\vecx) \equiv \phi(\vecx) \left( \partial_i \partial_j -\frac{1}{2} \partial^2 \delta_{ij} 
          \right) \phi(\vecx),
\label{qcontinuo}
\eeq
where $i,j=\{x,y\}$. This  symmetric and traceless tensor has, in two dimensions,  
only two independent elements, which essentially represent the mean orientation of domain 
walls and the strength of the orientational order. Similarly to the vector order parameter, a non zero value of $Q_{ij}$
represents a breakdown of the rotational symmetry $O(2)$. However, differently from a vector, this order parameter is 
invariant under rotations by $\pi$ characterizing a nematic symmetry.

A comprehensive description of the
physical meaning of this order parameter can be found in ref. \onlinecite{StBa2010}. Since the
tensor is symmetric, it can be diagonalized and in the principal axis base it has only one independent component, which can be
written as:
\beq
\langle Q \rangle = \int d^2k\,k^2\, \cos{(2\theta)} \, C(\veck),
\label{qcontinuoglobal}
\eeq
where $C(\veck)$ is the structure factor 
of the system and  $\vec k=k(\cos{\theta},\sin{\theta})$. This expression makes possible the 
computation of $\langle Q \rangle$ from experimental or numerical simulation data, since the structure factor is a usual observable. 
In this way, the orientational order parameter quantifies the 
degree of anisotropy of the domain pattern (i.\ e.\ the anisotropy in $C(\veck)$).

In ref. \cite{BaSt2007} it was shown that this
nematic-like order parameter emerges naturally from symmetry considerations in the
Landau expansion of the free energy of a system with isotropic competing interactions.
At mean field level plus Gaussian fluctuations~\cite{BaSt2009},  the phase transition
turns out to be in the Kosterlitz-Thouless universality class. This seems
to be a robust result, and was originally obtained in a somewhat different context by
Toner and Nelson~\cite{ToNe1981}.

Due to the ubiquity of orientational phases, intermediate between crystal and fluid ones,
in a great variety of condensed matter systems, it is important to have a unified microscopic description. 
This would also allow us to make closer contact with experiments and simulations.
With this motivation,  and also trying to clarify the nature 
of the nematic-like interaction term in the Ginsburg-Landau free energy analyzed in refs. 
\cite{BaSt2007,BaSt2009}, in this paper we develop a general classic statistical mechanics approach 
to obtain the phase diagram of generic competitive systems with Ising variables.

Consider the following Ising Hamiltonian on a square lattice:
\beq
{\cal H} = \frac{1}{2}\sum_{i,j} J_{ij}S_iS_j - \sum_i B_i S_i
\label{ising}
\eeq
where $\{S_i=\pm 1, i=1\ldots N\}$ are $N$ Ising variables, the sum runs
over all $i,j$ pairs in the lattice and $B_i$ is an external magnetic field. 
The interaction matrix $J_{ij}$ is
assumed to have a ferromagnetic part, $J_{ij}=J<0$ if $i,j$ are nearest-
neighbor sites (NNS) and an antiferromagnetic part $J_{ij}>0$ if $i,j$ are 
not NNS. These are the only constraints in order to apply the method. 

Consider a  discrete version  of the nematic order parameter defined in (\ref{qcontinuo}) coupled 
to a conjugate external nematic field through:
\beq
h_k^{ab} \left(\Delta_a S_k\Delta_b S_k-\frac{\delta_{ab}}{2} (\Delta S_k)^2 \right) 
\label{ncoupling}
\eeq
where $k$ is the lattice index, $a,b=x,y$ and $\Delta_a$ are lattice derivatives.
The field $h_k^{ab}$ has the same symmetries as the tensor order parameter, then
we can choose the coordinate system along the principal axes and write the conjugate 
field $h_k^{ab}$ as :
\beq
h_k^{ab}=\left(
\begin{array}{cc}
 h_k  &   0 \\
    & \\
  0  &   -h_k
\end{array}
\right)
\eeq
The coupling term of equation (\ref{ncoupling}) can be written as

\beq
\frac{1}{2} \sum_{ij} h_i \;  S_i K_{ij} S_j,
\label{nematicfieldenergy}
\eeq
where the matrix $K$ is given by
\beq
K_{ij}=\left\{ 
\begin{array}{lcl}
+1  & \mbox{if}  & j=i\pm \hat x \\
 & & \\
-1  & \mbox{if}  & j=i\pm \hat y
\end{array}
\right.
\eeq
$\hat x$ and $\hat y$ are unit vectors along the $x$ and $y$ axis of the square lattice.
In this way, the Hamiltonian now reads:

\beq
{\cal H} = \frac{1}{2}\sum_{i,j} \left(J_{ij}-h_i K_{ij}\right) S_iS_j - \sum_i B_i S_i
\label{ext-h}
\eeq

The local orientational order parameter is defined as :
\beq
\langle N_i\rangle=\frac{1}{\beta}\frac{\partial \ln Z}{\partial h_i}= \langle S_i S_{i+\hat x}-S_i S_{i+\hat y}\rangle ,
\eeq
where $Z = \sum_{\{S\}}\exp{\{-\beta {\cal H}\}}$ is the canonical partition function, with $\beta=1/k_BT$.
This order parameter has only one component and not two, like the one of eq. (\ref{qcontinuo}). If $\langle N_i\rangle$  is positive, the director points along
the $x$ direction wile if it is negative, the director mainly points in the $y$ direction. These are the only possible directions of the director. For this reason, if $\langle N_i\rangle\neq 0$, the resulting anisotropic phase is called Ising-Nematics, since it breaks the rotational point group of the lattice and it is invariant under rotations by $\pi$. Along the paper, for brevity,  we generally use the term ``nematic'' to refer to this phase, however, whenever we deal with a square lattice  model, ``Ising-Nematic''  should be understood.  

The global order parameter can be written as 
\beq
 Q = \frac{1}{2}\sum_{ij} K_{ij} \langle S_iS_j\rangle .
\label{globalnematicparameter}
\eeq
Equation (\ref{globalnematicparameter}) is completely analogous to the continuous version given by equation (\ref{qcontinuoglobal}). The anisotropic matrix $K$ for the Ising-Nematic phase, play the same role of $k^2\cos 2\theta$ for the nematic one. 
Thus, in the same way that eq. (\ref{qcontinuoglobal}) measures the degree of anisotropy of the structure factor $C(\veck)$, the 
order parameter (\ref{globalnematicparameter}) describes the degree of anisotropy
in the nearest-neighbor correlation functions between the $x$ and $y$ directions
of the lattice. A slightly different form of this order parameter has been
used before in simulations of stripe forming systems~\cite{BoMaWhDe1995,CaStTa2004,CaMiStTa2006}. 
In the present work, we attempt to compute it in a statistical mechanics framework. 

The technical problem of computing  the order parameter
reduces to the calculation of nearest-neighbor correlation functions in a proper approximation. In
this work, we will describe a mean-field like approximation 
based on the use of a Hubbard-Stratonovitch (HS) transformation.

Introducing a real variable on the lattice ($\Phi_i$, where $i$ is the lattice index) by means of a HS transformation on the original Hamiltonian (\ref{ising})(see, e.g. ref. \onlinecite{BiDoFiNe1995}) and exactly summing up the Ising variables $S_i$, 
we obtain an effective Hamiltonian given by
\begin{widetext}
\beq
{\cal H}_{eff}[\{\Phi\}] = 
\frac{1}{4}\sum_{ij}\Phi_iJ_{ij}\Phi_j - \frac{1}{2} \sum_i B_i \Phi_i - \frac{1}{\beta}
               \sum_i \log{\cosh{\left(\beta\sum_j J_{ij}\Phi_j\right)}}.
\label{effective}
\eeq
\end{widetext}
Using the general relation between correlations of the original discrete and continuous
variables~\cite{BiDoFiNe1995}:
\beq
\langle S_iS_j \rangle = -\frac{1}{2\beta} J_{ij}^{-1} +\frac{1}{4} \langle \Phi_i \Phi_j \rangle,
\label{spincorrel}
\eeq
for the cases of interest in this work, of an isotropic interaction matrix $J_{ij}$, the
order parameter reads:
\beq
 Q = \frac{1}{8}\sum_{ij} K_{ij} \langle \Phi_i\Phi_j \rangle
\label{barQ}
\eeq

\subsection{Mean field approximation for the order parameter $Q$}

To compute $Q$ given by equation (\ref{barQ}), 
we begin by considering the partition function 

\beq
Z={\cal N}\;\int {\cal D}\Phi\; e^{-\beta H_{eff}(\Phi)}
\label{pf}
\eeq
where ${\cal N}$ is a normalization constant and $H_{eff}(\Phi)$ is given by eq. (\ref{effective}).
We want to introduce an order parameter with nematic symmetry. For this purpose we introduce a symmetric 
traceless tensor $Q_k^{a,b}$, where $k$ is the lattice index and $a,b=x,y$ and write the partition function as 
\beq
Z={\cal N'}\;\int {\cal D}\Phi\; e^{-\beta H_{eff}(\Phi)}\int {\cal D}Q\; e^{-\beta \sum_k Tr (Q_k^2)}
\eeq
where ${\cal D}Q= \prod_k d Q^{a,b}_k$. Note that, in the square lattice, the introduction of a fully 
symmetric traceless tensor of rank two is redundant,
since the Ising-Nematic order parameter has only one independent component. However, we prefer to use this notation to stress that the method is general and can be also used to treat continuous systems with $O(2)$ symmetry, where the full tensor is needed to describe the phase transition. 
 
The Gaussian integral in $Q$ is simply a constant which can be  absorbed in the normalization factor 
${\cal N}'$. Shifting $Q_k\to Q_k-N_k(\Phi)$, keeping the measure ${\cal D}Q$ invariant we have,
\beq
Z={\cal N'}\;\int {\cal D}\Phi\; e^{-\beta H_{eff}(\Phi)}\int {\cal D}Q\; e^{-\beta \sum_k Tr (Q_k-N_k)^2},
\label{pfQ}
\eeq
where the symmetric traceless tensor $N_k(\Phi)$ will be defined later. 

At mean field level, $Q_k^{ab}$ is given by the saddle-point equation:
\beq
\frac{\partial\ln Z}{\partial Q_k^{ab}}=0,
\eeq
which reads
\beq
Q_k^{ab}= \langle N_k^{ab}\rangle .
\label{Qab}
\eeq
The average is given by:
\beq
\langle N_k^{ab}\rangle =\frac{1}{Z(Q)}\int {\cal D}\Phi\;  N_k^{ab}(\Phi)\; e^{-\beta\left ( H_{eff}(\Phi)+\sum_k Tr (Q_k-N_k)^2\right)}
\eeq

Choosing the coordinate system along the $Q_k^{ab}$ principal axes
\beq
Q_k^{ab}= Q_k\; \left(
\begin{array}{cc}
1 & 0 \\
0 & -1 
\end{array}.
\right)
\eeq
and
\beq
N_k^{ab}= N_k(\Phi)\; \left(
\begin{array}{cc}
1 & 0 \\
0 & -1 
\end{array}.
\right)
\eeq
and defining $\sum_k N_k=\frac{1}{8}\sum_{ij} K_{ij} \Phi_i\Phi_j$,
the effective Hamiltonian takes the form:  
\begin{widetext}
\begin{eqnarray}
H[\{\Phi_i,Q_i\}] &=& 
\frac{1}{4}\sum_{ij}\Phi_i J_{ij} \Phi_j -\frac{1}{\beta}
               \sum_i \log{\cosh{\left(\beta\sum_j J_{ij}\Phi_j\right)}}\nonumber \\
&+&  2 \sum_i  (N_i)^2 +2 \sum_i  Q_i^2
                -4 \sum_i  Q_i N_i,
\label{H}
\end{eqnarray}
\end{widetext}
where the external magnetic field $B_i$ has been set to zero.
This is a non-quadratic Hamiltonian in the variables $\Phi_i$, in the presence of 
a ``mean field'' orientational order parameter $Q$. The next step is to integrate the partition function in $\Phi_i$ considering 
quadratic fluctuations of $\Phi_i$ around a saddle-point approximation. 

Thus,  the expansion of $H$ up to quadratic order in the $\Phi$ fluctuations reads:

\beq
H(\Phi)=H(\Phi_{SP})+\frac{1}{2!}\sum_{ij} \left.H''_{ij}\right|_{\Phi=\Phi_{SP}} \delta\Phi_i\delta\Phi_j 
\eeq
where $\Phi_{SP}$ is the saddle-point solution $H'(\Phi_{SP})=0$, $H''_{ij}=\frac{\partial^2 H}{\partial 
\Phi_i\partial \Phi_j}$ and $\delta\Phi=\Phi-\Phi_{sp}$.

In this approximation $\langle \delta \Phi_i \delta \Phi_j\rangle=\frac{1}{4\beta} (H'')^{-1}_{ij}$
and then the mean field equation for $Q$, eq. (\ref{barQ}), reduces to:
\beq
 Q = \frac{1}{32\beta}\sum_{ij} K_{ij}(H'')^{-1}_{ij}.
\label{QH}
\eeq

In order to get the desired self-consistent equation for $Q$,
it is necessary to find an explicit expression for $H''(\Phi_{SP})$.
After a lengthy but straightforward calculation, we obtain for the Hessian:
\begin{widetext}
\beq
\left. H''_{lm} \right|_{SP}= \frac{1}{2}\, J_{lm} - \beta \sum_i 
             \frac{J_{il}J_{im}}{\cosh^2\left(\beta\sum_k J_{ik}\Phi_k^{SP}\right)}
       - Q K_{lm} + 
        \frac{1}{16} \sum_k \left(\Phi_k^2 \right)_{SP} K_{kl}K_{km}
\label{hessian2}
\eeq
\end{widetext}
Inserting this result in (\ref{QH}) a self-consistent equation for $Q$ is obtained.
It depends on the field $\Phi^{SP}$ and its square, which can be computed within the
same mean field approximation. 

The previous result allows in principle the calculation of positional ($\Phi$) and
orientational ($Q$) order parameters. 
We expect that a purely orientational phase, with $Q\neq 0$ and $\Phi^{SP}=0$,
which will be identified with a nematic phase with orientational order of domain walls,
but without positional order, may be present for a class of systems with competing
interactions. Considering this situation, the
simplest self-consistent approximation for the nematic order parameter reads, in
matrix notation:
\beq
Q=\frac{1}{16\beta}\  {\rm Tr}\left\{ \frac{K}{J-2\beta J^2-2 Q K}\right\}.
\label{Qmf}
\eeq
This self-consistent equation for $Q$ is the analog of the Curie-Weiss approximation
for the magnetization in the Ising model. 
This is one of the main results of our paper. If equation (\ref{Qmf}) has a non-trivial solution 
$Q\neq 0$, then the system exhibit an anisotropic but homogeneous phase with nematic symmetry. 
The presence, or not, of this phase depends of the detailed structure of the competing interactions, 
coded in the explicit form of the matrix $J$.

Guided by the results of the continuum Ginsburg-Landau model~\cite{BaSt2007}, we
search for a critical point signaling an isotropic-nematic transition:
Defining $A=J-2\beta J^2$ and expanding the r.h.s. of (\ref{Qmf}) in $ Q Tr(K/A) \ll 1$ we have 
\begin{widetext}
\beq
16\beta Q \approx Tr\left( \frac{K}{A} \right) +  2Q\, Tr \left( \frac{K}{A} \right)^2
         + 4Q^2\, Tr \left( \frac{K}{A} \right)^3 + 8Q^3 Tr \left( \frac{K}{A} \right)^4.
\eeq
\end{widetext}
If $A$ is rotationally invariant (invariant under discrete rotations in the square lattice), 
then $Tr \left( \frac{K}{A} \right) = Tr \left( \frac{K}{A} \right)^3= 0$. 
Then $Q=0$ is always a solution of the saddle point equations. If $Q\neq 0$:
\beq
\left[ 16\beta - 2 Tr \left( \frac{K}{A} \right)^2 \right]Q 
- 8Q^3 Tr \left( \frac{K}{A} \right)^4 =0, \hspace{2cm}
\eeq
and then, for $Q \ll1$:
\beq
Q \approx \frac{1}{2} \left[ \frac{8\beta - 
Tr \left( \frac{K}{A} \right)^2}{Tr \left( \frac{K}{A} \right)^4} \right]^{1/2}
\label{Qaprox}
\eeq
This result implies a continuous, second order {\em isotropic-nematic} transition, at a critical 
temperature given by:
\beq
\beta_c=\frac{1}{8}\  Tr \left(\frac{K}{A(\beta_c)}\right)^2.
\label{Tcritica}
\eeq
The existence of a solution of eq. (\ref{Tcritica}) depends on the detailed structure of the matrix $A$.
In the next section, we will show an explicit calculation on a model system in which this transition
is present.

\section{The dipolar frustrated Ising ferromagnet}
\label{dipolar}

The dipolar frustrated Ising ferromagnet  (DFIF) is a simple model for studying the thermodynamic
phases of ultrathin ferromagnetic films with strong perpendicular anisotropy
~\cite{DeMaWh2000,PoVaPe2003,WoWuCh2005,CaMiStTa2006,ViSaPoPePo2008}. 
In the strong anisotropy limit, the
anisotropic term of the classical dipole-dipole interactions is zero because the dipolar
moments point perpendicular to the lattice plane, and the system degrees of freedom are
well represented by Ising variables pointing perpendicular to the plane of the lattice. 
Furthermore, the strong uniaxial (perpendicular) anisotropy renders the energy symmetric
with respect to rotations on the lattice.
The Hamiltonian with competition between short-range exchange and long-range dipolar
interactions can be written as~\cite{CaMiStTa2006}:

\begin{equation}
{\cal H}= -\frac{J}{2} \sum_{<i,j>} S_i S_j + \frac{g}{2} \sum_{(i,j)}
\frac{S_i S_j}{r^3_{ij}}.
 \label{Hdipolar}
\end{equation}
\noindent The first sum runs over all pairs of nearest
neighbor spins, and the second one over all pairs of spins
of the lattice; $r_{ij}$ is the distance, measured in lattice
units, between sites $i$ and $j$. 
The relevant parameter is the ratio between the exchange $J>0$ and the dipolar $g>0$
intensities. Then, we fix $g=1$ without loosing generality. Note that the ferromagnetic,
short range exchange interaction is frustrated by the long range, antiferromagnetic,
dipolar interaction. The possibility of an Ising
nematic phase in these systems has been anticipated theoretically by Abanov et al.
~\cite{AbKaPoSa1995}, and numerical evidence from Monte Carlo simulations has been 
reported in \onlinecite{CaMiStTa2006}. Several experimental works have reported results 
which show domain patterns that could be identified with a nematic 
phase~\cite{PoVaPe2003,WoWuCh2005}, but up to now its characterization and properties have not been discussed.
Also, several 
theoretical, mainly numerical works have shown results of orientational order 
parameters~\cite{RoDe1990,HuSi1992,StSi2002}, but no quantitative characterization or 
distinction between, e.g. stripe, smectic and nematic phases have been attempted so far. 

\subsection{Nematic phase}

We have numerically solved  equation (\ref{Tcritica}) for the DFIF model (\ref{Hdipolar}),
in the thermodynamic limit where the  linear size of the system $L \to \infty$.
In reciprocal space, eq. (\ref{Tcritica}) reads:
\beq
\beta_c=\frac{1}{8}\  \int \ddk \left[ \frac{K(\veck)}{J(\veck)(1-2\beta_c J(\veck))}
                \right]^2.
\label{Tcont}
\eeq
where $J(\veck)$ and $K(\veck)$ are the Fourier transforms of the matrices $J_{ij}$ and $K_{ij}$.
The anisotropic function $K(\veck)$ in the square lattice is given by 
\beq
K(\veck) =2(\cos{k_x a}-\cos{k_y a})
\eeq
where $a$ is the lattice spacing. 
The Fourier transform of the interaction function is given by 
\beq
J(\veck) = -J\,L_{nn}(\veck)+g\,L_{dip}(\veck).  
\eeq 
where the the Fourier transform of the nearest-neighbor
interaction $L_{nn}(\veck)=2J(\cos{k_x a}+\cos{k_y a})$ and the dipolar interaction 
for small $k$  is approximated\cite{comment1} by $L_{dip}(\veck) \approx 1-k a+(ka)^2/4$

\begin{figure}[!htb]
\begin{center}
\includegraphics[scale=0.3,angle=-90]{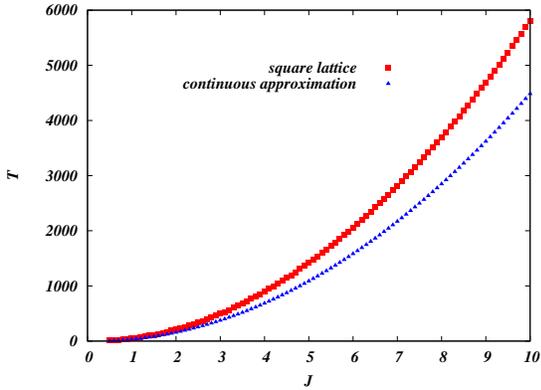}
\caption{(Color online) Isotropic-nematic transition line in the dipolar frustrated 
Ising ferromagnet. }
\label{dipolar_phase_diagram}
\end{center}
\end{figure}

We have solved equation (\ref{Tcont}) in two different limits:  keeping the full structure of the functions
$K(\veck)$ and $J(\veck)$ where the lattice symmetry is preserved,  and keeping only the leading order terms in  $ka\ll 1$,  where the effects of the lattice are suppressed (continuum approximation). 
The phase diagram in the $T,J$ plane is shown in figure \ref{dipolar_phase_diagram} for the two cases
considered. Both cases give similar results in this mean field approach.
The temperature scale depends on the value of the cutoff needed to get convergence of
the integrals. This is a well known limitation of the HS transformation~\cite{Amit1978} 
when the interaction matrix is not positive definite, as in this case. 
In any case, as the approximation is of a mean field
nature,  we only expect that the phase diagram be qualitatively 
correct.

An isotropic-nematic transition without positional order, $\langle \Phi \rangle =0$, is 
obtained in both cases. It is important to stress that in the full lattice
calculation, the low temperature phase is in fact an Ising-nematic, while in the continuum 
limit the phase is a nematic one. This last case corresponds to the result found 
in ref.\onlinecite{BaSt2007} from a phenomenological Ginsburg-Landau model. 
A fundamental difference between both cases arise in the nature of the transitions. 
While in the continuum case the
transition is of the KT type, in the first case the universality class is probably
Ising (see ref. \onlinecite{AbKaPoSa1995} for a discussion of different scenarios when lattice
anisotropy is considered), but this remains to be proved for this model. 
The present microscopic approach for the nematic phase has the Ginsburg-Landau Hamiltonian studied in 
refs. \cite{BaSt2007,BaSt2009} as the continuum limit. Note that the continuum limit of
equation (\ref{Qmf}) corresponds to equation $(12)$ of ref. \onlinecite{BaSt2007}.

Interestingly, orientational order develops at higher temperatures for systems where the
competing interaction is weak, i.e. for weak frustration. This corresponds to the
experimental situation in ultrathin ferromagnetic films, where visual inspection of
the domains formed point to the existence of a nematic phase~\cite{PoVaPe2003,
SaRaViPe2010}.

\begin{figure}[!htb]
\begin{center}
\includegraphics[scale=0.3,angle=-90]{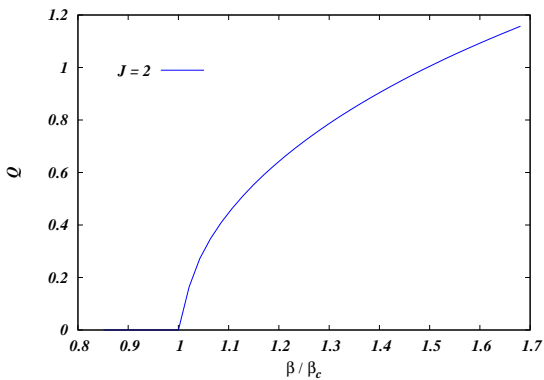}
\caption{(Color online) Mean field nematic order parameter for $J=2$.}
\label{op}
\end{center}
\end{figure}

In figure \ref{op} we show the order parameter $Q$ near the transition, obtained
solving eq. (\ref{Qaprox}) for the lattice case.
It shows the second order nature of the phase transition. This is an expected result since, in two dimensions, the nematic symmetry only allows even powers of the order parameter (nematic as well as Ising-Nematics) in the free energy\cite{chaikin-1995}. This is different from three dimensions where the transition is of first order.  However, as was already indicated,  fluctuations could probably drive the transition to a different class. 

\subsection{Positional order}

At lower temperatures, besides the orientational order represented by the nematic phase,
positional order may also emerge. Within the present framework, the positional order
parameter is given by the density field $\Phi_i$, which can be computed together
with $Q(T)$ from equations (\ref{H}) and (\ref{hessian2}). We have not pursued this
in the present work, which is devoted to the description of the nematic phase. Instead,
it is interesting to analyze the behavior of the two-point spin correlation function
in the nematic phase, which gives information of the possible growth of positional
order at high temperatures.

\begin{figure}[!htb]
\begin{center}
\includegraphics[scale=0.5,angle=0]{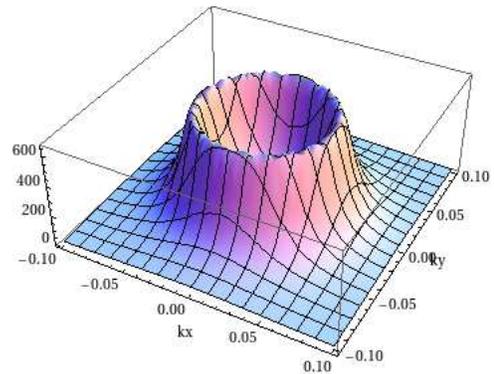}
\caption{(Color online) Structure factor of the DFIF near the paramagnetic-stripes
phase transition, without considering the nematic correction term, computed from
eq. (\ref{structure}) for J=10.}
\label{structure_isotropic}
\end{center}
\end{figure}

\begin{figure}[!htb]
\begin{center}
\includegraphics[scale=0.5,angle=0]{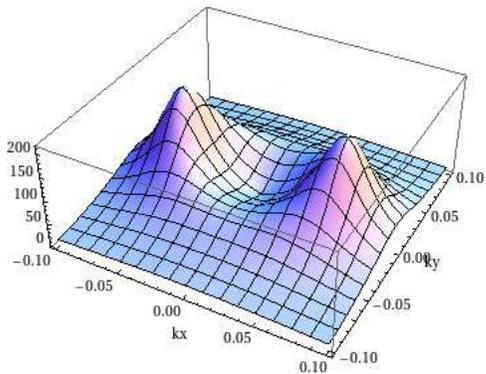}
\caption{(Color online) Structure factor of the DFIF in the nematic phase, 
near the nematic-stripes phase transition for J=10.}
\label{structure_nematic}
\end{center}
\end{figure}

The spin-spin correlation
function, eq. (\ref{spincorrel}), in reciprocal space reads:
\beq 
G(\vec k) =  \frac{1}{J(\vec k)-2\beta J(\vec k)^2-2 Q K(\vec k)}-\frac{1}{J(\vec k)} 
\label{structure}
\eeq
This equation must be solved self-consistently with eq. (\ref{Qmf}). 
If the term proportional to $Q(T)$ in the denominator is disregarded, we end with
the usual, mean field approach to the computation of modulated phases.
In figure \ref{structure_isotropic} we show the
function $G(\vec k)$ for $Q=0$, for a temperature slightly above the mean-field critical
temperature, which in this case is given by $T_c = max_{\vec k} J(\vec k)$. The
high temperature profile of $G(\vec k)$ is isotropic and a phase transition to a
modulated phase with characteristic wave vector $k_0 \neq 0$ takes place through
a breaking of a continuous rotational symmetry. This is the ``Brazovskii's scenario''
for the isotropic-stripes phase transition in systems with nearly isotropic 
competing interactions~\cite{Br1975}.
Instead, upon inclusion of the nematic order parameter correction, the spectrum above
the low temperature modulated phases changes in an essential way. In fact, 
because the nematic transition breaks rotational symmetry, the structure factor is
anisotropic in this phase, as shown in figure \ref{structure_nematic}, again for a
characteristic temperature just above the transition to a modulated phase with
positional order (divergence of a staggered magnetic susceptibility).
Note that the spectrum has a broad support in the plane of wave vectors, a clear
indication that there is no positional order in the nematic phase. Nevertheless,
it shows two well defined maxima at characteristic wave vectors $\pm \vec k_0$, 
which show the $\pi$-rotation symmetry of the nematic phase. 
Furthermore, both peaks are on the x-axis,
a consequence of the breaking of continuous symmetry already imposed by the square
lattice. This is the structural signature of an Ising-nematic phase. 

Another interesting question regards the (in)commensurability of the characteristic
wave vectors. In this solution, as expressed e.g. by equation (\ref{structure}), 
it is clear that the wave vector which maximizes the structure factor in the
nematic phase {\em depends continuously on temperature}. Then, at least in the
nematic phase, the characteristic wave vectors are incommensurate with the lattice.

\begin{figure}[!htb]
\begin{center}
\includegraphics[height=.2\textheight,width=.4\textwidth,angle=0]{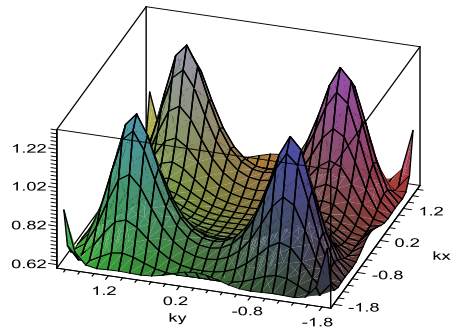}
\caption{(Color online) Structure factor of the BNNNI model near the paramagnetic-stripes
phase transition, computed from
eq. (\ref{structure}) and (\ref{JBNNNI}). $\delta=1.8$ and $\beta=1/13$.}
\label{BNNNI}
\end{center}
\end{figure}

A very different situation arises in short-ranged interaction models like anisotropic next-nearest-neighbor Ising (ANNNI)\cite{VillBak1981} or the  biaxially next-nearest-neighbor Ising  model (BNNNI)\cite{AyYa1989}. This class of models could be similarly treated within our formalism. For instance,  the Fourier transform of the interaction matrix $J_{ij}$ for the BNNNI model reads, 
\begin{eqnarray}
J(k)&=& 2 \delta \left(\cos(k_x a)+\cos(k_y a)\right) \nonumber \\
&-&2\left(\cos(2 k_x a)+\cos(2 k_y a)\right)
\label{JBNNNI}
\end{eqnarray}
where $a$ is the lattice constant and $\delta$ measures the competition between first and second neighbors interactions.
In figure (\ref{BNNNI}) we show the structure factor of this model, computed form  eq. (\ref{structure}) in the high temperature phase.
We observe the presence of four peaks obeying the lattice symmetry. The peaks weights grow as the temperature is lowered , signaling the onset of positional order. However, the position of the peaks in reciprocal space are given by $\cos(k_x a)=\cos(k_y a)=\delta/2$ which do not depend on temperature. This fact allows a possible incommensurate as well as a commensurate positional order at low temperatures. Concerning the main focus of this paper (the nematic phase), we have computed the self-consistent equation (\ref{Tcritica})  for the nematic critical temperature and we have found no solution for this case. This is consistent with the general belief that it is necessary to have a macroscopically degenerate number of ordering wave vectors to produce a pure orientational ordered phase (compare fig. (\ref{structure_isotropic}) with fig. (\ref{BNNNI})).

\section{Conclusions}
\label{conclusions}

We have developed an approach to study phase transitions in two dimensional Ising
systems with competing interactions which may show orientational and positional order.
A nematic order parameter in the square lattice has been defined, suitable to quantify 
the degree of orientational  order of interfaces, useful in systems which
show microphase separation. A mean field approach has been developed which leads 
to a set of self-consistent equations for orientational and positional order
parameters, similar in spirit to the Bragg-Williams approximation. The approach is
very general, and can be applied in principle to any Ising system with competing
interactions. Nematic, smectic and stripe-crystal phases can be studied within this
framework. We solved the self-consistent equations for the Dipolar Frustrated Ising
Ferromagnet. This model for ultrathin ferromagnetic films with perpendicular anisotropy
is known to have a striped low temperature phase and the ground state is striped for
arbitrary small dipolar interaction. Within the present approach we have gone beyond
the usual mean field approximation for the stripe phase, showing that an isotropic-
nematic phase transition takes place at a higher temperature than the mean field
isotropic-stripe transition. 

Comparing with experimental evidence, as discussed in
the Introduction, a nematic phase
without positional order is probably present in ultrathin ferromagnetic films, although
it has not been characterized already~\cite{PoVaPe2003,WoWuCh2005}. Our results indicate
that the nematic phase is more robust for higher values of the ratio between the 
exchange and dipolar interactions, $J/g$, as is the case in experiments where the
intensity of the dipolar term is two or three orders of magnitude weaker than the 
exchange interaction.

Comparison with numerical results from computer simulations
is still difficult for several reasons. The first one is that our calculation is of 
mean field character and then our results can only be qualitatively correct. A
second problem is that up to now there have been very few attempts to quantify and
characterize orientational order in systems of the type considered in this work, 
even in computer simulations~\cite{RoDe1990,HuSi1992,StSi2002,CaStTa2004,NiSt2007}
To our knowledge, the most detailed simulational study of the phase transitions in
the DFIF presented so far is the work
by Cannas et. al.~\cite{CaMiStTa2006}, where evidence was shown of an intermediate
nematic phase, between a paramagnetic and a stripe-crystal phase. It that
reference, small values of the ratio $J/g$ were considered. In fact, the nematic phase
was reported for a ratio $J/g=2$, which was the larger value considered in that work.
For that case, the nematic phase was observed in a narrow temperature interval, in 
qualitative agreement with our phase diagram of figure \ref{dipolar_phase_diagram}.
Note also that for $J/g=1$ no evidence of nematic phase was reported in that work,
again in agreement with our results, which point to the absence of nematic phase for
$J\leq1$. It it still a major challenge for computer simulations to attain ratios in
the experimental range $J/g \sim 10^2-10^3$. To our knowledge, the largest values
attained in simulations have been around $J/g \sim 10$, which can only yield a narrow
nematic phase~\cite{StSi2002,NiSt2007}. 

As already said, the present approach is valid for arbitrary microscopic interactions, which
enter only in the computation of the final self-consistent equations. Because of its
generality, we expect it can be useful for facing some important yet unsolved
problems regarding the behavior of systems with competing interactions. One of these
problems is the role of the range of interactions in producing pure orientational
phases. The necessity of long range interactions is frequently invoked, but the actual
influence of the relative range of the competing interactions is still an open
problem. 

Some points
have to be addressed in order to turn the approach quantitative. The computation of
the nematic order parameter is equivalent to the computation of nearest-neighbors
correlation functions. Then, better approximations for the computation of correlation
functions will turn the results quantitatively reliable. Another point is the formal
problem with the non-positive character of the quadratic form, necessary for applying
the Hubbard-Stratonovich transformation. Some previous works have already addressed
this question~\cite{GlGrKaKoSaZi2007,PaFi1999}, which will be a subject of future work.

Although we have not
tried to get quantitative results, we hope the present approach will be
useful to describe in full  the phase transitions in two dimensional systems 
which show microphase separation and nematic-like orientational phases originated from 
competing interactions.

This work was partially supported by CNPq and FAPERJ (Brazil). We wish to thank
Sergio Cannas and Alessandro Vindigni for a critical reading of the manuscript.


%

\end{document}